\documentclass[usenatbib]{mn2e}
\usepackage[english]{babel}
\usepackage[T1]{fontenc}
\usepackage{amsmath}                                                            
\usepackage{graphicx}

\usepackage{aecompl}

\title[Spiral arm kinematics: Gaia Mock Observations]
{The stellar kinematics of co-rotating spiral arms in Gaia mock observations}
\author[J. A. S. Hunt et al.]
  {\parbox{\textwidth}{Jason A. S. Hunt$^{1}$\thanks{E-mail: jason.hunt.11@ucl.ac.uk}, Daisuke Kawata$^{1}$, Robert J. J. Grand$^{2,3}$, Ivan Minchev$^4$, \\      Stefano Pasetto$^{1}$ and Mark Cropper$^{1}$}\vspace{0.5cm}
\\
$^{1}$ Mullard Space Science Laboratory, University College London,
Holmbury St. Mary, Dorking, Surrey, RH5 6NT, UK \\
$^{2}$ Heidelberger Institut fur Theoretische Studien, Schloss-Wolfsbrunnenweg 35, 69118 Heidelberg, Germany \\
$^3$ Zentrum f$\ddot{u}$r Astronomie der Universit$\ddot{a}$t Heidelberg, Astronomisches Recheninstitut, M$\ddot{o}$nchhofstr. 12-14, 69120 Heidelberg, Germany \\
$^4$ Leibniz-Institut f$\ddot{u}$r Astrophysik Potsdam (AIP), An der Sternwarte 16, 14482 Potsdam, Germany \\
}
\date{Submitted to MNRAS: $22^{nd}$ December 2014.} 

\pagerange{\pageref{firstpage}--\pageref{lastpage}}
\pubyear{2014}

\begin{document}

\maketitle

\label{firstpage}

\begin{abstract}
We have observed an $N$-body/Smoothed Particle Hydrodynamics simulation of a Milky Way like barred spiral galaxy. We present a simple method that samples $N$-body model particles into mock $Gaia$ stellar observations and takes into account stellar populations, dust extinction and $Gaia's$ science performance estimates. We examine the kinematics around a nearby spiral arm at a similar position to the Perseus arm at three lines of sight in the disc plane; $(l,b)=(90,0), (120,0)$ and $(150,0)$ degrees. We find that the structure of the peculiar kinematics around the co-rotating spiral arm, which is found in \cite{KHGPC14}, is still visible in the observational data expected to be produced by $Gaia$ despite the dust extinction and expected observational errors of $Gaia$. These observable kinematic signatures will enable testing whether the Perseus arm of the Milky Way is similar to the co-rotating spiral arms commonly seen in $N$-body simulations.
\end{abstract}

\begin{keywords}
methods: $N$-body simulations --- methods: numerical --- galaxies: structure
--- galaxies: kinematics and dynamics --- The Galaxy: structure
\end{keywords}

\section{Introduction}
\label{intro-sec}
The spiral features visible in many galaxies have long been the subject of debate. Although it has been almost a century since the resolution of the ``great debate'' of \cite{SC21}, when it was argued over whether these beautiful spiral structures were nebulae within our galaxy or galaxies in their own right, the mechanisms which generate them are still uncertain. One of the problems with developing a comprehensive theory of spiral arms is the so called ``winding dilemma". It is known from observations of disc galaxies that the stars in the inner region have a higher angular velocity than those in the outer region. Therefore the spiral structure should ``wind up" relatively quickly if the spiral arms rotate at the mean rotation velocity of the stars \citep[e.g.][]{W96}, contrary to observations of many ``grand design'' spiral galaxies. A proposed solution to the winding dilemma is given by spiral density wave theory \citep{LS64} which treats the spiral structure as a density wave which can rotate rigidly as a feature with a constant pattern speed and thus be long lived.

However, no $N$-body simulations have yet been able to reproduce these long lived stable spiral arms, despite the increase in computational power and resolution which has occurred in recent years \citep[e.g.][]{S11,DB14}. Recent work has shown spiral modes and waves which survive over multiple rotations \citep{QDBMC11,RDL13,SC14} while the spiral arm features in the stellar mass are short-lived but recurrent \citep[e.g.][]{SC84,CF85,B03,FBSMKW11,GKC12,GKC12-2,GKC13,BSW13,RFetal13,DVH13} including in galaxies with a central bar \citep[e.g.][]{GKC12-2}, implying that the large spiral arms visible in external galaxies may only $appear$ to be rigid structures extending over the disc, while in fact being made of transient reforming features. 

The interpretation of the transient and recurrent spiral arm features observed in $N$-body simulations is still in debate. For example, \cite{MFQDCVEB12} showed for the first time (by studying the time evolution of the disc power spectrum) that spiral wave modes in $N$-body simulations can last for as long as 1 Gyr, which can justify treating the wave modes as quasi-stationary structure, and the transient and recurrent spiral arm features can be explained by the superposition of different modes with different pattern speeds \citep[see also][]{RDQW12,SC14}. On the other hand, \cite{GKC12,DVH13,BSW13} demonstrated non-linear growth of the spiral arm features due to similar but different (in terms of evolution) mechanisms from swing-amplification \citep{T81}, which could be difficult to explain with the linear superposition of the wave modes.

Our position within the Milky Way gives us a unique view of these spiral structures seen in external galaxies, but it comes with its own set of problems which we must overcome when studying them. The location and kinematics of the gaseous component of the arms may be determined from HI and CO observations \citep[e.g.][]{DHT01,NS03,KK09}. However to observe the kinematics of the stellar component in and around the spiral arms we must look through the disc plane, which carries the heaviest levels of dust and gas, and thus high levels of extinction. 

Dust extinction has long been a problem for Milky Way model construction. Although there are reasonably reliable extinction maps for extra galactic sources whose extinction by the interstellar medium of the Milky Way can be corrected as a function $A_\lambda(l,b)$ \citep[e.g.][]{SFD98}, three dimensional extinction mapping for sources within the Milky Way i.e. $A_\lambda(l,b,d)$ is more challenging. There are three dimensional extinction maps for individual sections of the sky \citep[e.g.][]{DS01,MRRSP06,HBJ14,SM14} and two dimensional maps have been extended to three dimensions \citep[e.g.][]{DCL03}. However a truly Galactic 3D extinction map does not yet exist \citep{RB13}. The European Space Agency (ESA)'s $Gaia$ mission will help us map the stellar structure and kinematics of the Milky Way, and help constrain extinction at the same time \citep{BJetal13}.

$Gaia$, which was launched on the 19th December 2013 will provide detailed astrometric \citep[e.g.][]{LLHOBH12}, spectroscopic \citep[e.g.][]{Ketal11} and photometric \citep[e.g.][]{Jetal10} information for around one billion stars in the Milky Way. Detailed information on $Gaia$ scientific accuracies is available in, for example, \cite{dB12}. Synthetic $Gaia$ mock data have already been used to demonstrate different applications of the real $Gaia$ data set. For example, \cite{AMAFR14} use three tracer populations (OB, A and Red Clump stars) with the $Gaia$ selection function, errors and dust extinction, and demonstrated that the $Gaia$ mock data can recover the parameters of the Galactic warp. \cite{RGFAAA14} examine the Galactic bar in the $Gaia$ observable space using Red Clump tracers with the $Gaia$ selection function, errors and dust extinction combined with selected Red Clump stars from the Apache Point Observatory Galactic Evolution Experiment \citep[APOGEE DR10, e.g.][]{Aetal14} showing the value of combining data from complimentary surveys. In \cite{HK13} we show that we can recover the large scale structure of the Galactic disc with our Made-to-Measure Galaxy modelling code, \sc{primal }\rm \citep{HKM13,HK12,HK13}, and make a good estimation of the patten speed of the bar, using tracer populations of M0III and Red Clump stars with the $Gaia$ selection function, errors and dust extinction.

There exist full mock catalogues of $Gaia$ stars, e.g. the $Gaia$ Universe Model Snapshot (\sc{gums}\rm) which provides a view of the Besan\c{c}on Galaxy model as seen from $Gaia$ \citep{Rea12}, taking into account dust extinction while assuming there are no observational errors. This detailed prediction of $Gaia$ observations gives an excellent indication of the volume and quality of data which will become available from $Gaia$, predicting 1.1 billion observable stars, almost 10,000 times more than from its predecessor $Hipparcos$. \sc{gums }\rm can be extended through the $Gaia$ Object Generator (\sc{gog}\rm) \citep{Xetal14} to simulate intermediate and final catalogue data including the introduction of realistic astrometric, photometric and spectroscopic observational errors to the catalogue based upon $Gaia$ science performance estimates. While these mock data provide an excellent example of the capabilities of $Gaia$, the Besan\c{c}on galaxy model is an axisymmetric model and a kinematic model not a dynamical model. Although $Gaia$ will not provide accelerations, the kinematics it will provide are from a dynamical system, the Milky Way. Thus it is important for our purpose to generate catalogues from fully dynamical models with non-axisymmetric structures, such as spiral arms and a bar, which for example $N$-body disc galaxy models can provide.

Therefore we propose here to create mock $Gaia$ observations from an $N$-body model using a population synthesis code such as \sc{galaxia }\rm \citep{SBJB11}, or the methodology presented in \cite{PCK12} or \cite{LWCKHFC14}. \sc{galaxia }\rm is a flexible population synthesis code for generating a synthetic stellar catalogue from an $N$-body or an analytical galaxy model over wide sections of the sky, with a sampling scheme which generates a smoothly distributed sample of stars. Synthetic catalogues generated from dynamical Galaxy models are essential for preparing to exploit the real $Gaia$ catalogue and can be used to determine whether certain features within the Milky Way will be visible to $Gaia$.

In our previous work \citep{KHGPC14} we examined the kinematics of both the stellar and gas components around a transient, co-rotating spiral arm in a simulated barred spiral galaxy similar in size to the Milky Way. Although this arm is transient, similar arms recur during the evolution of the galaxy. We made predictions of observable kinematic signatures that may be visible in the Milky Way's Perseus arm if it is also a transient, recurrent and co-rotating spiral arm. We then compared our simulation with data from APOGEE and the maser sources from \cite{Retal14} measured by the Bar and Spiral Structure Legacy (BeSSeL) survey and the Japanese VLBI Exploration of Radio Astronomy (VERA), finding tentative agreement between our simulation and the observations. Owing to the low number of maser sources and the lack of distance information for the APOGEE stars no firm conclusions could be drawn; however it is encouraging to see similar features in both, including the possible signatures of a co-rotating spiral arm. 

In this paper we build upon the previous work by generating a stellar sample with different populations from the simulation data in \citet{KHGPC14} and making mock observations of these stars taking into account the expected $Gaia$ science performance estimates. The aim is not to make further predictions about the kinematics of transient, recurrent and co-rotating spiral arms but rather to examine whether these signatures, remain visible in the $Gaia$ data if they exist in the Milky Way.


\section{Simulation}
\label{sim}

We use the simulated galaxy which is presented in \citet{KHGPC14} and \cite{GKC14b}. The details of the numerical simulation code, and the galaxy model are described in \cite{KHGPC14}. We briefly describe the galaxy model in this section. The galaxy is set up in isolated conditions, and consists of a gas and stellar disc but no bulge component. The discs are embedded in a static dark matter halo potential \citep{RK12,KHGPC14}. The dark matter halo mass is $M_{\rm dm}=2.5 \times 10^{12}$ $\rm M_{\odot}$, and the dark matter density follows the density profile from \cite{NFW97}, with a concentration parameter of $c=10$. The stellar disc is assumed to follow an exponential surface density profile with the initial mass of $M_{\rm d,*} = 4.0 \times 10^{10}$ $\rm M_{\odot}$, a radial scale length of $R_{\rm d,*} = 2.5$ kpc and a scale height of $z_{\rm d,*} = 350$ pc. The gas disc is set up following the method of \citet{SDMH05}, and has an exponential surface density profile with the scale length of $R_{d,g} = 8.0$ kpc. The total mass of the gas is $10^{10}$ $\rm M_{\odot}$. The simulation comprises $10^6$ gas particles and $4 \times 10^6$ star particles; therefore each particle has a mass of $10^4$ $\rm M_{\odot}$. The resolution is sufficient to minimise numerical heating from Poisson noise \citep{FBSMKW11,S13}. We employ a minimum softening length of $158$ pc (equivalent to a Plummer softening length of $53$ pc) with the spline softening and variable softening length for gas particles as suggested by \citet{PM07}. 

The radial profile of the mean metallicity of stars and gas is initially set by $\mathrm{[Fe/H]} (R) = 0.2 - 0.05(R/1 \text{ kpc})$,
and the metallicity distribution function at each radius is centred on the mean metallicity value with the dispersion set to a Gaussian distribution of $0.05$ dex for the gas and $0.2$ dex for the stars. The stellar ages are set randomly between 0 and 10 Gyr for stars present at the beginning of the simulation.

\begin{figure}
\centering
\includegraphics[width=\hsize]{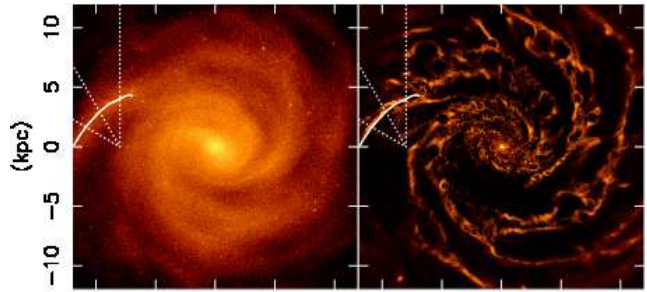}
\caption{Snapshot of the simulated galaxy from \citet{KHGPC14} which is also used in this paper. The left (right) panel shows the face-on view of the star (gas) particle distribution. The solid line indicates the position of the spiral arm identified. The observer is assumed to be located at $(x,y)=(-8, 0)$ kpc. Three line-of-sight directions ($l_{\rm LOS}=90, 120$ and 150 deg) are highlighted with the dotted lines. The galaxy is rotating clockwise.}
\label{galaxy}
\end{figure}

The simulation was run for 1 Gyr from the initial conditions with the $N$-body smoothed particle hydrodynamics code, \sc{gcd+ }\rm \citep[e.g.][]{KG03,RK12,BKW12,KOGBC13,KGBGR14} without the inclusion of any continuous external inflow of gas for simplicity. In this paper we use the same snapshot of the galaxy as used in \cite{KHGPC14} which is taken at $t=0.925$ Gyr, as this snapshot shows a spiral arm at a similar location to the Perseus arm from the Milky Way (see Fig. \ref{galaxy}).

\section{$Gaia$ mock catalogue}
\label{Model}

\label{Kawata14}
In \cite{KHGPC14} the kinematics of the spiral arm shown in Fig. \ref{galaxy} are examined at three lines of sight $l_{\rm LOS}=90, 120$ and 150 deg, with $b_{\text{los}}=0$ because of the lower extinction relative to other lines of sight in the plane. Predictions are made of the observational signatures of co-rotating spiral arms notably the difference in kinematic structure between the trailing near side and leading far side of the spiral arm. In general, in \cite{KHGPC14} (as also shown in \cite{GKC14}) the stars in the trailing near side rotate slower because they tend to be at the apo-centre and migrate outward, and the stars in the leading far side rotate faster as they tend to be at the peri-centre and migrate inward. There are however some stars which follow the opposite trend, leading to multiple populations seen in the rotational velocity in the leading far side; one faster, and one slower than the single population in the trailing near side. These features which will be discussed later may be caused by the co-rotation resonance of the spiral arm, and are visible at different galactic longitudes because the arm in the simulation co-rotates at all the examined radial range. However, in \cite{KHGPC14}, the spiral arm kinematics are examined using the full, error and extinction free $N$-body data and thus such trends, when present, are easy to identify. 

In this Section we describe how we generate a sample of stars from the $N$-body model of \cite{KHGPC14} to produce a mock $Gaia$ catalogue. It is worth noting that the population synthesis code, \sc{galaxia }\rm \citep{SBJB11} provides a tool to generate stellar populations from $N$-body simulation data. However, because we plan to combine such a tool with our Made-to-Measure Galaxy modelling code, \sc{primal}\rm, we have developed our own simplified version of \sc{galaxia}\rm, a population synthesis code called \sc{snapdragons}\rm, (Stellar Numbers And Parameters Determined Routinely And Generated Observing $N$-body Systems). \sc{snapdragons }\rm uses the same isochrones and extinction map as \sc{galaxia}\rm, but uses a different and more simplistic process to generate the stellar catalogue which is described in Section \ref{Synth}. \sc{snapdragons }\rm allows us to add the expected Gaia errors more easily, and enables us to track the link between sampled stars and their parent $N$-body particle for our future studies, e.g. \sc{primal }\rm modelling of the Galactic disc by fitting tracers from multiple stellar populations, and identifying radially migrating stars and non-migrating stars trapped by the spiral arm \citep{GKC14}.

\subsection{Extinction}
\label{Ex}
We use the extinction map of the Milky Way taken from \sc{galaxia }\rm \citep{SBJB11}, which is a 3D polar logarithmic grid of the dust extinction constructed using the method presented in \citet{BKF10} and the dust maps from \citet{SFD98}. The same extinction is applied in \cite{HK13} and more detail is given there. In an update from \citet{HK13} we follow the correction to the Schlegel $E_{B-V}$ presented in \citet{Setal14} such that
\begin{equation}
E_{B-V}=E_{B-V}\biggl(0.6+0.2\biggl(1-\text{tanh}\biggl(\frac{E_{B-V}-0.15}{0.1}\biggr)\biggr)\biggr).
\end{equation}
This correction is made as it has been suggested \citep[e.g.][]{AG99,YFS07} that the reddening is overestimated by the maps from \cite{SFD98} by $\sim$1.3-1.5 in regions with high extinction with $A_V>0.5$ ($E_{B-V}>0.15$). This correction reduces extinction by $\sim40\%$ for low latitude high extinction regions but has minimal effect on high latitude low extinction regions.

\subsection{Population Synthesis: \sc{snapdragons}\rm}
\label{Synth}

The goal of this population synthesis code is to split each $N$-body particle from the galaxy simulation into an appropriate number of stellar particles creating a mock catalogue of observable stars from our $N$-body model. We must choose an IMF and a set of isochrones with which to work. We choose a Salpeter IMF \citep{S55} where the IMF, $\Phi(m)$, is defined in each mass interval d$m$ as
\begin{equation}
\Phi(m)\text{d}m=Am^{-(x+1)}\text{d}m,
\end{equation}
where $x=1.35$ is the Salpeter index, and $A$ is a constant for normalisation in the desired mass range. We set this constant as
\begin{equation}
A_i=m_i\biggl(\int_{m_{\star,\text{min}}}^{m_{\star,i,\text{max}}}m^{-x}\text{d}m\biggr)^{-1},
\end{equation}
where $m_i$ is the $N$-body particle mass, $m_{\star,i,\text{max}}$ is the maximum initial mass of any surviving star and $m_{\star,\text{min}}$ is the minimum stellar mass to be considered. We make use of the Padova isochrones \citep[e.g.][]{BBCFN94,MGBGSG08}, although the choice of isochrones (and IMF) may be substituted with others with no change to the methodology. 

It is worth noting that the Padova isochrones are available only for stellar masses above 0.15 $M_{\odot}$. \sc{galaxia }\rm for example uses the isochrones from \citet{CBAH00} to extend the mass limit down to 0.07 $M_{\odot}$, which is the hydrogen mass burning limit. We set our lower limit on stellar mass as $m_{\star,\text{min}}=0.1$ $M_{\odot}$ to correspond with the simulation from \cite{KHGPC14} and extrapolate from the Padova isochrones for $0.1\leq M_{\odot}\leq0.15$. It is relatively safe to do this because all such stars lie on the main sequence. Additionally these exceedingly faint stars will not be visible at the distance of the spiral arms which are the focus of this work.

As discussed in Section \ref{sim} each $N$-body star particle in the simulated galaxy has been assigned an age and metallicity within the chemodynamical code \sc{gcd+}\rm, then it is made to evolve. When we examine the snapshot, each particle is matched to its nearest isochrone in both metallicity and age from the grid of isochrones which are extracted from \sc{galaxia}\rm. Once an isochrone is selected, we identify $m_{\star,i,\text{max}}$ from the isochrone. We then determine how many stars to sample from the $N$-body particle by integrating the IMF over the desired mass range;
\begin{equation}
N_s=A\int_{m_{\star,i,<V_{\text{lim}}}}^{m_{\star,i,\text{max}}}m^{-(x+1)}\text{d}m,
\end{equation}
where $m_{\star,i,<V_{\text{lim}}}$ is minimum mass required for the star particle to be brighter than our apparent magnitude selection limit, $V_{\text{lim}}$, taking into account the extinction value at the position of the parent particle. Stars smaller than $m_{\star,i,<V_{\text{lim}}}$ are not used in the subsequent analysis, to save on computational time.

We then randomly sample stellar masses from the section of the isochrone $N_s$ times. We have weighted the random selection by the IMF using the equation
\begin{equation}
m_{\star}=(R m_{\star,i,\text{max}}^{-x}+(1-R)m_{\star,i,<V_{\text{lim}}}^{-x})^{\frac{1}{-x}},
\end{equation}
where $R$ is a random number between 0 and 1.

The isochrones are comprised of discrete stellar data, and therefore we then interpolate within the nearest isochrone values of $M_V$ and $V-I_c$ to determine $M_{V_{\star}}$ and $V-I_{c\star}$ for the generated $m_{\star}$. At this stage we assume the generated stars have the same position and velocity as their parent particles.

\subsection{Observational Errors}
\label{Error}
Having generated the visible stellar catalogue we then add observational errors based upon the $Gaia$ Science Performance estimates\footnote{http://www.cosmos.esa.int/web/Gaia/science-performance}. We use the post launch error estimates approximated from the estimates in pre-launch performance by Merc\`{e} Romero-G\'{o}mez \citep[e.g.][]{RGFAAA14}, provided through the Gaia Challenge collaboration\footnote{http://astrowiki.ph.surrey.ac.uk/dokuwiki/doku.php}. We assume the position and velocity of the Sun is known. We locate the observer at ($-8$,0,0) kpc as shown in Fig. \ref{galaxy}, and the motion of the Sun is assumed to be 228 km s$^{-1}$. For this work, while generating the stellar catalogue we produced stars only brighter than $V_{\text{lim}}\leq16$ mag, which is well within $Gaia's$ $m_G\leq20$ mag magnitude limit for the astrometry. However, because we are interested in the Galactic radial and rotation velocity for the stars, which requires the full 6D phase space information, we chose the lower magnitude limit where $Gaia$ RVS can produce the reasonably accurate line-of-sight velocity. Note that the errors are added to the parallax, proper motion and line-of-sight velocities. 

A full description of the method to add the pre-launch $Gaia$ error is available in \cite{HK13}. However the $Gaia$ science performance estimates have been revised after launch, and as such a correction must be made. The error in parallax has increased, and although it has little effect for stars with $m_V\leq16$ mag which we work with in this paper, the coefficients within the equation to describe the pre-launch parallax performance (provided by Kazi, Antoja \& DeBruijne (Oct. 2014) by fitting to the new estimations on the $Gaia$ science performance web page) are revised to
\begin{eqnarray}
\sigma_{\pi}&=&(-11.5+706.1z+32.6z^2)^{1/2} \nonumber \\ & &\times(0.986+(1-0.986)(V-I_c)),
\label{sigpi}
\end{eqnarray}
where
\begin{equation}
z=\text{max}(10^{0.4(12-15)},10^{0.4(G-15)}),
\label{zmax}
\end{equation}
correcting also the typo for equations (\ref{sigpi}) and (\ref{zmax}) in \cite{HK13}.

Additionally, because of the loss of spectroscopic accuracy by $\sim1.5$ mag in the RVS post launch performance we also apply a correction to the error function for the end of mission radial velocity. We change the table\footnote{http://www.cosmos.esa.int/web/Gaia/table-5} of values $a$ and $b$, again determined by fitting the revised performance estimates on the $Gaia$ science performance web page, for the equation
\begin{equation} 
\sigma_{v_r} = 1 + b\text{e}^{a(V-14)},
\end{equation}
where $a$ and $b$ are constants dependant on the spectral type of the star. The new table along with the code to add the $Gaia$ error is available online\footnote{https://github.com/mromerog/Gaia-errors}.


\section{Results}
\label{R}
As discussed in Section \ref{Kawata14}, it was shown in \cite{KHGPC14} that in general the stars in the trailing near side of the spiral arm rotate slower than average because they tend to be at the apo-centre, and the stars in the leading far side of the spiral arm rotate faster than average as they tend to be at the peri-centre. However, there are groups of stars which follow different trends leading to multiple populations which will be discussed later. It is important to determine whether such features will still be visible in the $Gaia$ catalogue, not just the error and extinction-free $N$-body model. In this Section we show the result of sampling these $N$-body data into stellar data, first looking at the properties of the resulting mock stellar catalogue, and then examining the spiral arm kinematics with the stellar data taking into account dust extinction and $Gaia$ science performance estimates.

\subsection{Population synthesis}
\begin{figure}
\centering
\includegraphics[width=\hsize]{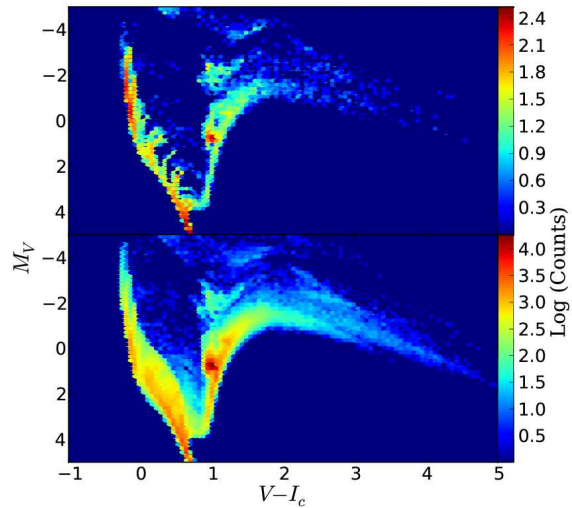}
\caption{Colour magnitude diagram for stars generated by \sc{snapdragons }\rm from particles within a square region of $\pm2$ deg (upper) and $\pm5$ deg (lower) around $(l,b) = (90,0)$ deg. Stars with apparent magnitude of $m_V\leq16$ only are included.}
\label{CMD}
\end{figure}

In this section we describe the stellar catalogue produced by \sc{snapdragons}\rm, and show the resulting colour magnitude diagram (CMD) varying the area of the sky coverage. Fig. \ref{CMD} shows the CMD for stars generated by \sc{snapdragons }\rm from particles within a square region of $\pm2$ deg (upper) and $\pm5$ deg (lower) around $(l,b) = (90,0)$ deg. The upper panel of Fig. \ref{CMD} shows clearly the individual stellar isochrones because there are only a small number of $N$-body particles in the selected region, and each particle has only one age and metallicity. These problems are resolved when smoothing is applied in the phase space distribution and age-metallicity distribution \citep[e.g.][]{SBJB11}. However, as discussed in Section \ref{Synth} we deliberately avoid this smoothing to maintain the clear particle-star relation. The lower panel of Fig. \ref{CMD} shows no such discrete structure, as there are sufficiently many particles to cover a broad range of stellar ages and metallicities in the CMD. Therefore, care is required with the resolution of the $N$-body simulation and the selection function if we discuss in detail the stellar population distribution in the CMD. However, this is unlikely to affect the study in this paper.

\subsection{Observable Spiral Arm Kinematics}
In this section we examine if the possible kinematic signatures of co-rotating transient and recurrent spiral arms identified in \cite{KHGPC14} will be visible in the $Gaia$ data even given the dust extinction in the disc and $Gaia's$ science performance accuracy. A detailed analysis of the kinematics themselves is the focus of \cite{KHGPC14}, while this work is concerned with the visibility of this kinematic structure in the $Gaia$ data. We examine the rotational velocities of the stars in the catalogue for different distances because in \cite{KHGPC14} we found the rotation velocity is most affected by the transient co-rotating spiral arm. Then we calculated the Probability Density Function (PDF) of the rotation velocity of stars behind and in front of the spiral arm using Kernel Density Estimation (KDE) which we are using as a desirable alternative to histograms \citep[e.g.][]{W06}. 

\begin{figure*}
\centering
\includegraphics[width=\hsize]{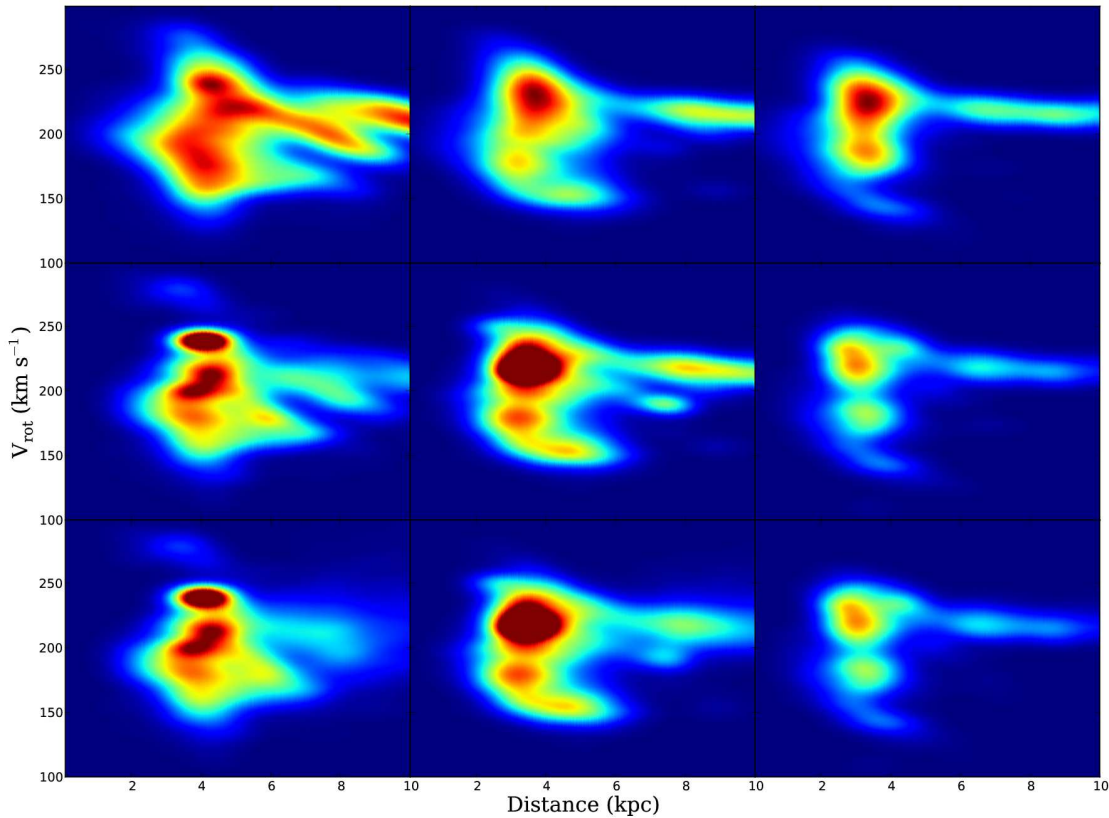}
\caption{Smoothed linear scale contour plot of galactocentric rotation velocity of simulation particles (upper), selected \sc{snapdragons }\rm stars (middle) and selected \sc{snapdragons }\rm stars observed with $Gaia$ error (lower) for $(l,b)=(90,0)$ (left), $(l,b)=(120,0)$ (middle) and $(l,b)=(150,0)$ (right). For the \sc{snapdragons }\rm stars (middle and lower panels), a limited selection of $M_V\leq-1$ calculated using $m_V$ and  $d_{\text{obs}}$ and assuming a known extinction, along with $\sigma_{v_r}/(v_r\times d_{\text{obs}})\leq0.15$ is shown to avoid overly dense populations of fainter stars at smaller distances. This is to visualise the data set, and these faint stars contribute to the subsequent analysis. Note while consistent for \sc{snapdragons }\rm stars across the different lines of sight, the density scale is different for the simulation particles; however the choice of the scale is arbitrary.}
\label{Crot}
\end{figure*}

Fig. \ref{Crot} shows a smoothed contour plot of the galactocentric rotational velocity against distance for particles and stars within a square region of $\pm5$ degrees around $(l,b)=(90,0)$ (left), $(l,b)=(120,0)$ (middle) and $(l,b)=(150,0)$ (right). This compares the kinematics of the underlying $N$-body model (upper) with the stellar catalogue generated with \sc{snapdragons}\rm, before (middle) and after (lower) the addition of the errors from the $Gaia$ science performance estimates. Owing to the high percentage of low mass and luminosity stellar types which would dominate the selected region and saturate the plot at small distances, we have made cuts to our sample to visualise the underlying kinematic structure from the stellar catalogue. We have first cut the sample of stars in all three lines of sight with absolute magnitude, $M_V\leq-1$, calculated from the apparent magnitude $m_V$ and observed distance $d_{\text{obs}}$, assuming the dust extinction is known. We then cut with $\sigma_{v_{\text{los}}}/(v_{\text{los}}\times d_{\text{obs}})\leq0.015$ kpc$^{-1}$ to select the stars with lower error in the line of sight velocities at a smaller distance to generate similar quantities of data at different distance scales. This is purely for illustration purposes and we are not suggesting that this is the best possible selection function. The upper panels of Fig. \ref{Crot} show the different kinematic structure in the $N$-body model at the different lines of sight. These are the same data as those shown in the top panels of Fig. 4 from \cite{KHGPC14}. Note that the density colour scale for the $N$-body data is different from the stellar data in the middle and lower panels. 

The middle row of panels of Fig. \ref{Crot} show the velocities of the selected stars, which appear slightly different from those of the $N$-body data owing to the selection function. While the general shape of the distribution has been recovered, at $(l,b)=(90,0)$ deg (middle left) the fast rotating stars within the arm dominate the density scale and wash out the rest of the plot slightly. At $(l,b)=(120,0)$ deg (middle), although there is some saturation around $220$ km s$^{-1}$ the kinematic structure is clearly visible and is a good match to the particle data. Similarly at $(l,b)=(150,0)$ deg (middle right), despite the lower number of counts, the kinematic structure is clearly shown. 

The lower panels of Fig. \ref{Crot} show the error affected rotation velocity and distance for the selected stars taking $Gaia$ science performance estimates into account. The rotation velocity is calculated from the observed parallax, proper motion and line of sight velocities. At $(l,b)=(90,0)$ (lower left) the shape of the distribution remains relatively unchanged, with the main loss in accuracy occurring around $d_{\text{obs}}\approx7-10$ kpc. The recovery of the kinematic structure around the spiral arm around $d_{\text{obs}}\approx4$ kpc remains almost identical to the case without observational errors. At $(l,b)=(120,0)$ (lower middle) the visible loss of accuracy is again in the outer region of $d_{\text{obs}}\approx7-10$ kpc, with the region containing the spiral arm remaining very similar to that of the error free case. At $(l,b)=(150,0)$ (lower right), the entire distribution remains very similar to the middle right panel, the case without $Gaia$ like observational errors. 

\begin{figure*}
\centering
\includegraphics[width=\hsize]{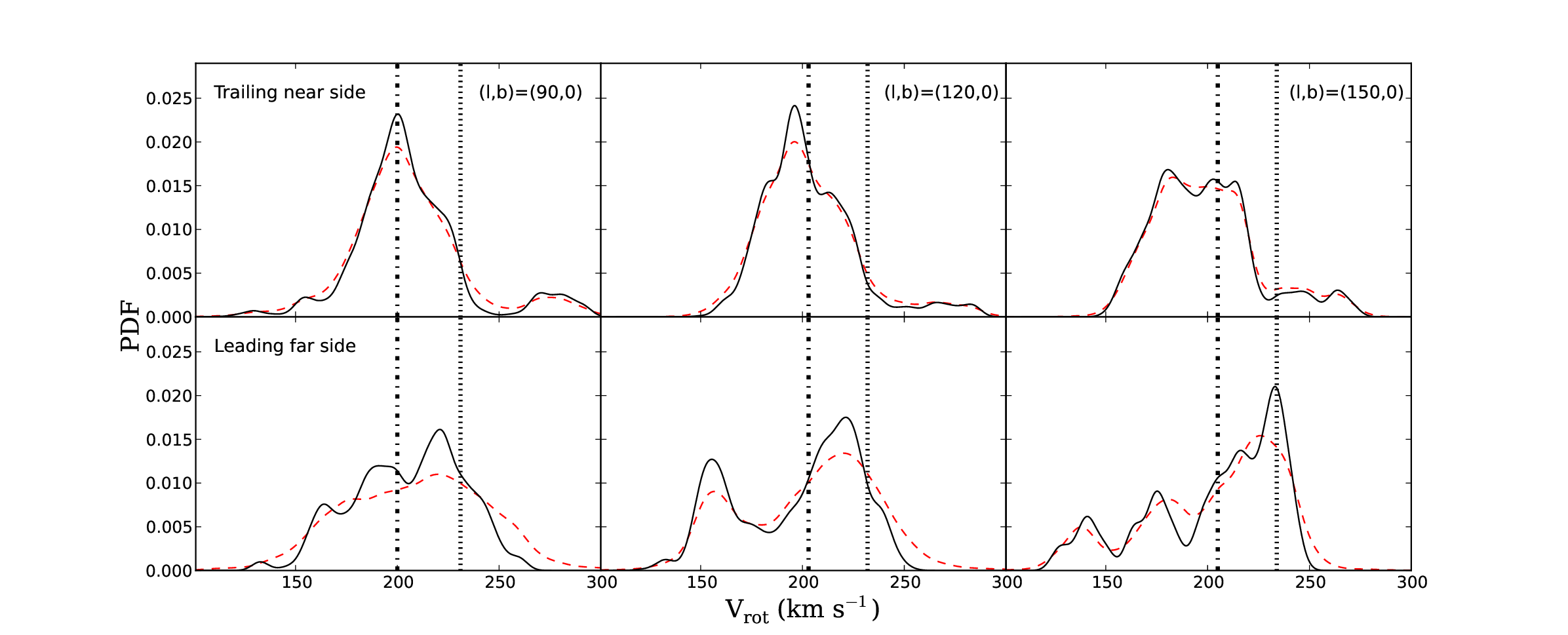}
\caption{Comparison of the distribution of galactocentric rotational velocities for the stars generated by \sc{snapdragons }\rm within a square region of $\pm5$ degrees around $(l,b)=(90,0)$ (left), $(l,b)=(120,0)$ (middle) and $(l,b)=(150,0)$ (right) in the trailing near side (upper) and leading far side (lower) of the spiral arm which meet the $m_V\leq16$ selection limit. The black solid curve shows the true velocities, and the red dashed curve shows the distribution once the Gaia errors have been applied. The vertical lines show the circular velocity (dotted) and the mean rotation velocity (dash-dotted) at the radius of the spiral arm.}
\label{Hrot}
\end{figure*}

Fig. \ref{Hrot} shows the PDF's, with a KDE bandwidth of 4, for the rotational velocity of the stars in the catalogue within a square region of $\pm5$ degrees around $(l,b)=(90,0)$ (left), $(l,b)=(120,0)$ (middle) and $(l,b)=(150,0)$ (right) in the trailing near side, between 1 and 2 kpc closer than the centre of the arm (upper) and leading far side between 1 and 2 kpc further than the centre of the arm (lower). Note that these distance bins were chosen as they show the discussed structure most clearly; the same features are present closer to the arm but are less clear. The centre of the arm was determined to be at $d=4.0$ kpc at $(l,b)=(90,0)$, $d=3.4$ kpc at $(l,b)=(120,0)$ and $d=3.3$ kpc at $(l,b)=(150,0)$. Note that Fig. \ref{Hrot} uses all the stars with $m_V\leq16$, not applying the selection function used for illustration purposes in Fig. \ref{Crot}. At all three lines of sight Fig. \ref{Hrot} shows a clear difference in the distribution of velocities for the `true' data (black solid) when comparing the different observed distances, as shown in \cite{KHGPC14}. This is a positive outcome considering the loss of data from the dust extinction. When comparing the `true' (black solid) stellar catalogue data with the stellar data taking into account dust extinction and $Gaia$'s expected errors (red dashed) a general smoothing out of the structure is evident in the `observed' data. The upper panels of Fig. \ref{Hrot} showing the trailing near side of the arm show very similar PDF's when comparing the true and observed stellar data, whereas the lower panels showing the leading far side show an information loss, especially at $(l,b)=(90,0)$, where the three peaks are no longer resolved. This is to be expected because of the higher distances and therefore additional extinction; however at $(l,b)=(120,0)$ and $(150,0)$ even on the far side of the spiral arm the structure within the distribution is still clearly visible.  

When comparing the `observed' data in Fig. \ref{Hrot} in front and behind the spiral arms, we see a clear difference in the PDF at all three lines of sight. In each case, the PDF in the trailing near side of the spiral arm forms a single central peak similar to the mean rotation velocity, with a small tail towards faster rotation velocities whereas the leading far side of the spiral arm shows a broader distribution of velocities with a peak velocity faster than the peak for the trailing near side. The difference is particularly apparent at $(l,b)=(120,0)$ deg where the leading far side shows two clear peaks, one faster and one slower than the single peak in the trailing near side. This bimodal distribution can also be seen in the lower middle panel of Fig. \ref{Crot} between 4.39 and 5.39 kpc (although note that Fig. \ref{Crot} uses a different selection function). Also at $(l,b)=(150,0)$ deg the single broad peak in the trailing near side is easily distinguishable from the leading far side which shows three peaks. These three peaks are also partially visible in the lower right panel of Fig. \ref{Crot} between 4.29 and 5.29 kpc. These features all match those observed in \cite{KHGPC14} despite the addition of dust extinction and observational errors to the data.

In general, as shown in \cite{GKC14}, the stars in the leading side rotate faster as they tend to be at peri-centre phase and migrating inward, and stars in the trailing side rotate slower as they tend to be at apo-centre phase and migrating outward. This explains the single large peak in the trailing side, and the largest peak on the leading side which has a higher rotational velocity than the single peak on the trailing side as shown in Fig. \ref{Hrot}. However, when the transient spiral arm starts forming, stars which are close to the arm on the trailing side and are close to the peri-centre phase, are accelerated towards the arm, passing through and then slowing down as they reach the apo-centre on the leading side as discussed in \cite{KHGPC14}. These stars correspond to the `slower' peaks visible in the lower panels of Fig. \ref{Hrot}. Similarly, the stars which are close to the arm and close to the apo-centre phase on the leading side are decelerated by the arm, and are overtaken by the arm. Then they are accelerated again by the arm once they are on the trailing side at peri-centre phase, which corresponds to the small tail present at high velocities in the upper panels of Fig. \ref{Hrot}. The difference in the rotation velocity distribution between the leading and trailing side of the spiral arm seen in Figs. \ref{Crot} and \ref{Hrot} is that the latter population is smaller than the former. It appears that it is easier for stars to escape from the arm on the leading side than the trailing side. From our analysis of $N$-body simulations this appears to be a common feature of transient and co-rotating spiral arms. 

\cite{CQ12} propose that the radial overlap of multiple longer-lived patterns moving at different pattern speeds can reproduce the transient spiral features, which when strong enough can lead to radial migration away from the co-rotation radius associated with co-rotating spiral arms as seen, for example in \cite{GKC12,GKC12-2}. In such a scenario, the spiral arm features are co-rotating, which may give rise to the co-existence of many inner and outer Lindblad resonances in a range of radii and lead to the features visible in Figs. \ref{Crot} and \ref{Hrot}. However, further analysis of the spiral arms in $N$-body simulations is required before drawing firm conclusions on the mechanism that generates such kinematic signatures, which we will tackle in future studies.

From Figs. \ref{Crot} and \ref{Hrot} we find that $Gaia's$ scientific accuracy ought to be sufficient to examine the kinematic structure of the nearby spiral arms in the Milky Way, even on the far side of the arm. Fig. \ref{Hrot} shows clear differences in the kinematics in the leading and trailing sides of the spiral arm, notably the difference in the number and locations of the peaks, and the small high velocity tail present in the trailing near side. The comparison between the middle and lower panels of Fig. \ref{Crot} shows little difference, implying that the observational error from $Gaia$ will have limited effect on our ability to study the kinematics of the spiral arms. Further examination of galaxy models constructed using the different theories of spiral arm formation will be essential to determine the distinct kinematic signatures of each theory.


\section{Summary}
\label{SF}
We observed our $N$-body/SPH simulation of a Milky Way like barred spiral galaxy to create a mock $Gaia$ stellar catalogue, with particular interest in the stellar kinematics in and around the spiral arms. We focused on the same three lines of sight in the disc plane as \cite{KHGPC14}, $(l,b)=(90,0), (120,0)$ and $(150,0)$ deg and analysed the galactocentric rotational and line of sight velocities of the selected stars as a function of the distance from the observer. In agreement with existing literature on $N$-body spiral galaxy simulations the spiral arm features seen in the stellar mass in our model are transient, recurrent and co-rotating, i.e. the spiral arm is rotating at the circular velocity of the stars at the selected lines of sight.

We show that the structure in the kinematics identified in \cite{KHGPC14} remains visible after the inclusion of dust extinction and observational errors based upon $Gaia$ science performance estimates. Although the inclusion of these observational effects makes the trends less clear, they are still observable in the mock $Gaia$ data in front of, inside and behind the spiral arm. The structure on the trailing near side is relatively unchanged, whereas the structure on the leading far side is, unsurprisingly, more affected, although the bi-modal (or more) and broader distribution of the rotation velocities is still clearly visible. Because we believe that these kinematic signatures are indications of transient and co-rotating spiral arms owing to the co-rotation resonance at all radii, we predict they should be visible in the $Gaia$ data at different longitudes if the Milky Way's Perseus arm is also a transient and co-rotating spiral arm.

Encouraged by the success of this study, we intend to repeat the analysis with simulated galaxies which use different theories of spiral structure formation, for example test particle simulations \citep[e.g.][]{MQ08,MBSB10,MF10,FSF14,Aetal14-2} and $N$-body simulations with a fixed spiral arm potential \citep[e.g.][]{WBS11}. From these analyses we expect to make predictions of the kinematic signatures of different spiral arm theories, which can be tested by the $Gaia$ stellar catalogue.

\section*{Acknowledgements}
We gratefully acknowledge the support of the UK's Science \& Technology Facilities Council (STFC Grant ST/H00260X/1 and ST/J500914/1). The calculations for this paper were performed on Cray XT4 at Center for Computational Astrophysics, CfCA, of the National Astronomical Observatory of Japan and the DiRAC facilities (through the COSMOS consortium) including the COSMOS Shared Memory system at DAMTP, University of Cambridge operated on behalf of the STFC DiRAC HPC Facility. This equipment is funded by BIS National E-infrastructure capital grant ST/J005673/1 and STFC grants ST/H008586/1, ST/K00333X/1 \& ST/J001341/1. The authors acknowledge the use of the IRIDIS High Performance Computing Facility, and associated support services at the University of Southampton. We would also like to thank PRACE for the use of the Cartesius facility. This work was carried out, in part, through the $Gaia$ Research for European Astronomy Training (GREAT-ITN) network. The research leading to these results has received funding from the European Union Seventh Framework Programme ([FP7/2007-2013] under grant agreement number 264895). We would also like to thank Merc\`{e} Romero-G\'{o}mez and Francesca Figueras for providing the subroutine to calculate the $Gaia$ performance errors, including the update to post launch estimates, Sanjib Sharma for providing the \sc{galaxia }\rm extinction maps and isochrones and Sami Niemi for the suggestion of using KDE's to visualise the velocity distributions.

\bibliographystyle{mn2e}
\bibliography{ref2}

\label{lastpage}
\end{document}